\documentclass{article}

\usepackage{PRIMEarxiv}

\usepackage[utf8]{inputenc} 
\usepackage[T1]{fontenc}    
\usepackage{hyperref}       
\usepackage{url}            
\usepackage{booktabs}       
\usepackage{amsfonts}       
\usepackage{nicefrac}       
\usepackage{microtype}      
\usepackage{fancyhdr}       
\usepackage{graphicx}       
\usepackage{doi}            
\usepackage{orcidlink} 

\usepackage{epsfig}
\usepackage{subcaption}
\usepackage{calc}
\usepackage{amssymb}
\usepackage{amstext}
\usepackage{amsmath}
\usepackage{amsthm}
\usepackage{multicol}
\usepackage{pslatex}
\usepackage{apalike} 
\usepackage{algorithm2e}
\usepackage{xcolor}
\usepackage{siunitx} 

\title{A Comparative Analysis of Statistical and Machine Learning Models for Outlier Detection in Bitcoin Limit Order Books
}

\author{
  Ivan Letteri \orcidlink{0000-0002-3843-386X} \\
  Department of Life, Health and Environmental Sciences \\
  University of L'Aquila, P.le S. Tommasi \\
  Coppito - 67100, L'Aquila, Italy\\
  \texttt{ivan.letteri@univaq.it} \\
}

\begin{document}
\maketitle

\begin{abstract}
The detection of outliers within cryptocurrency limit order books (LOBs) is of paramount importance for comprehending market dynamics, particularly in highly volatile and nascent regulatory environments. This study conducts a comprehensive comparative analysis of robust statistical methods and advanced machine learning techniques for real-time anomaly identification in cryptocurrency LOBs. Within a unified testing environment, named AITA Order Book Signal (AITA-OBS), we evaluate the efficacy of thirteen diverse models to identify which approaches are most suitable for detecting potentially manipulative trading behaviours. An empirical evaluation, conducted via backtesting on a dataset of 26,204 records from a major exchange, demonstrates that the top-performing model, Empirical Covariance (EC), achieves a 6.70\% gain, significantly outperforming a standard Buy-and-Hold benchmark. These findings underscore the effectiveness of outlier-driven strategies and provide insights into the trade-offs between model complexity, trade frequency, and performance. This study contributes to the growing corpus of research on cryptocurrency market microstructure by furnishing a rigorous benchmark of anomaly detection models and highlighting their potential for augmenting algorithmic trading and risk management.
\end{abstract}


\section{Introduction}
\label{sec:introduction}

The cryptocurrency market is characterised by extreme volatility, pronounced liquidity fluctuations, and a comparatively underdeveloped market microstructure relative to traditional financial markets \cite{KOUTMOS2018122}, such as equities and bonds. These distinctive characteristics render the market particularly susceptible to manipulative trading behaviours, including spoofing, wash trading, and layering, which can distort the price discovery process and undermine market integrity \cite{fratric_manipulation_2022}. Despite these challenges, the decentralised nature of cryptocurrencies like Bitcoin has garnered significant interest from both retail and institutional investors, catalysing the demand for sophisticated tools to detect and mitigate market anomalies in real time.

Recent scholarship has underscored the pivotal role of order flow—the continuous stream of buy and sell orders—in determining price movements and liquidity within cryptocurrency markets \cite{APERGIS2021101685,DIMPFL2021100584}. However, prevailing approaches to outlier detection in limit order books (LOBs) often fail to adequately capture the high-frequency, non-linear dynamics inherent in these markets, resulting in suboptimal performance under extreme conditions. This study addresses this lacuna by conducting a rigorous comparative analysis of a wide array of statistical and machine learning models. To ensure a fair and reproducible comparison, we developed the AITA Order Book Signal (AITA-OBS), a unified testing environment designed to evaluate each model's performance under identical conditions. AITA-OBS is a module of the main application called Artificial Intelligence Trading Assistant (AITA) \cite{AITAletteri23}, which integrates the module Volatility Trading System (VolTS) \cite{letteri2023volts} used in combination with AITA-OBS. 

Our research builds upon a growing body of literature in cryptocurrency market microstructure, extending the findings of studies such as \cite{Kercheval2015Yuan}, who applied support vector machines (SVMs) to predict mid-price movements in LOBs, and \cite{Tadi_2021}, who demonstrated the efficacy of cointegration-based trading strategies in cryptocurrency markets. By systematically evaluating models ranging from Empirical Covariance and Histogram-Based Outlier Score (HBOS) to One-Class SVM and CBLOF, this work provides a comprehensive performance benchmark.

The primary contributions of this study are threefold: (i) the development of a unified and transparent backtesting environment (AITA-OBS) for evaluating anomaly detection models in LOBs; (ii) a rigorous comparative analysis of thirteen statistical and machine learning models for outlier detection; and (iii) an empirical evaluation, including transaction cost analysis, that demonstrates the potential of outlier-driven strategies to enhance algorithmic trading and portfolio risk management. Our results indicate that the top-performing model outperforms the Buy-and-Hold (B\&H) benchmark by 6.70\%, highlighting the value of this approach for navigating the complexities of modern cryptocurrency markets.

This paper is structured as follows: Section \ref{sec:background} provides essential background and a review of related work. Section \ref{sec:methodology} details the entire experimental design, including the dataset, feature engineering, the mathematical formulation of the tested models, and the pipeline for converting outlier scores into trading signals. Section \ref{sec:results} presents the backtesting results, a comparative performance analysis, and a discussion of the findings. Finally, Section \ref{sec:conclusion} concludes the paper with a summary of our contributions and avenues for future research.

\section{Background and Related Work}
\label{sec:background}

\subsection{A Taxonomy of Outliers in Limit Order Books}
\label{subsec:outlier_taxonomy}

\begin{itemize}
    \item \textbf{Volume Outliers:} Orders with volumes significantly deviating from the typical distribution.
    \begin{itemize}
        \item \textit{High-Volume Orders:} A buy order with a volume substantially above the historical average may induce a temporary price spike, potentially signalling the activity of a 'whale' (an investor with significant capital) \cite{chohan_bitcoins_2017}.
        \item \textit{Low-Volume Orders:} Orders with exceptionally small volumes may be indicative of manipulative practices, such as placing phantom orders to distort the perception of market depth.
    \end{itemize}
    \item \textbf{Temporal Outliers:} Irregularities in the timing of order arrivals, such as anomalous intervals between order timestamps. We compute the \textit{Time Gap}; an absence of new orders for an extended period in an otherwise liquid market may signal a technical issue or an unforeseen market event. Temporal outliers can have cascading effects on spreads and market depth.
    \item \textbf{Liquidity Outliers:} Anomalous changes in the depth of the order book.
    \begin{itemize}
        \item \textit{Depth Decline:} A sudden withdrawal of orders from the book (liquidity withdrawal) may signal manipulative strategies like layering.
        \item \textit{Artificial Depth Increase:} The placement of a large volume of non-bona fide orders to influence the price without the intention of execution \cite{WangW17}.
    \end{itemize}
    \item \textbf{Volatility Outliers:} Extreme price fluctuations over very short timeframes.
    \begin{itemize}
        \item \textit{Sudden Volatility Shock:} An abrupt and transient change in price, often caused by large volume orders or external events such as news flashes.
        \item \textit{Anomalous Calm:} Prolonged periods without significant price changes in a typically volatile market may indicate manipulative activity.
    \end{itemize}
\end{itemize}

\subsection{Related Work}
Market microstructure is the field of study concerned with the processes governing price formation, liquidity, and efficiency in financial markets. Studies by Apergis \cite{APERGIS2021101685} and Dimpfl et al. \cite{DIMPFL2021100584} have highlighted the complexities of price discovery and liquidity provision in cryptocurrency markets, emphasising their divergence from traditional financial instruments. While these studies provide foundational insights, they often lack robust implementations for anomaly detection within high-frequency trading contexts. This study extends these findings by integrating advanced statistical and machine learning models specifically tailored to the high-frequency dynamics of cryptocurrency LOBs.

In the context of trading strategies, Tadi et al. \cite{Tadi_2021} evaluated a dynamic cointegration-based pairs trading strategy on the BitMEX exchange. Their results demonstrated that, despite the limitations of the market microstructure, their approach outperformed a naive buy-and-hold (B\&H) strategy, showcasing the potential for exploiting market inefficiencies through statistical techniques.

Traditional statistical models have long been applied to address challenges in market microstructure. For instance, \cite{math9010056} compared market liquidity predictions between crypto and fiat currencies using ARMA and GARCH time series models, alongside the K-Nearest Neighbours (KNN) approach. They found that ARMA performed well for developed fiat markets, while GARCH was better suited for emerging market currencies, highlighting the importance of selecting models appropriate to the market's characteristics.

In parallel with these developments, machine learning has emerged as a powerful tool for analysing and predicting financial market dynamics, particularly in the context of high-frequency LOB data. Kercheval and Yuan \cite{Kercheval2015Yuan} proposed a support vector machine (SVM) framework to predict mid-price movements and spread crossings in LOBs.

These studies provide a foundation for our work, which is inspired by the aforementioned research but focuses on a synergistic combination of statistical models and machine learning techniques for detecting and managing outliers in Bitcoin order flow.

\section{Experimental Design and Methodology}
\label{sec:methodology}

\subsection{Dataset and Feature Engineering}\label{sect:dataset}
\begin{figure}[!ht]
	\centering
	\includegraphics[width=35em]{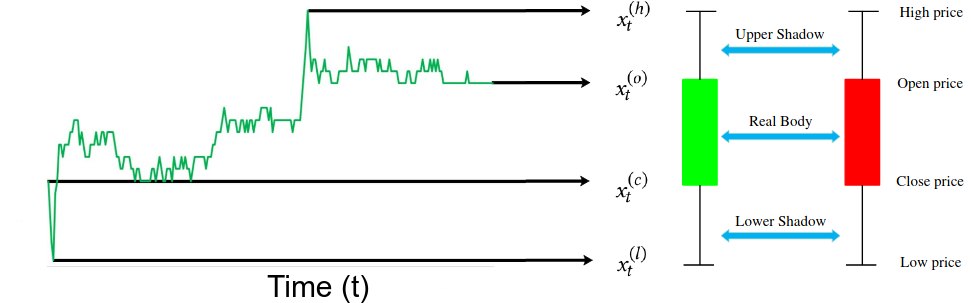}
	\caption{An example of a candlestick chart.}
	\label{fig:candlestick}
\end{figure}

AITA-OBS is a module integrated within the AITA framework \cite{letteri_trading_2023} \cite{letteri2022dnnforwardtesting}, where Price Action (PA) is encoded as OHLC (Open, High, Low, Close) price data, as represented in candlestick charts (see Figure \ref{fig:candlestick}). For each timeframe $t$, the OHLC of an asset is represented as a 4-dimensional vector $X_t = (x^{(o)}_t,x^{(h)}_t,x^{(l)}_t,x^{(c)}_t)^T$, accompanied by the trading volume $v_t$. The constraints are $x_t^{(l)} > 0$, $x_t^{(l)} < x_t^{(h)}$, and $x_t^{(o)}, x_t^{(c)} \in [x_t^{(l)},x_t^{(h)}]$.

The features engineered for outlier detection are derived from OHLC data and the LOB layers (bid and ask) to capture key market dynamics and microstructure properties:

\begin{itemize}
    \item \textit{Execution Price ($p_t$):} The transaction price at time $t$. Large deviations from the expected price identify outliers, measured using the standard deviation $\sigma(p_t)$ over short intervals.
    \item \textit{Bid and Ask Prices ($b_t, a_t$):} The highest bid price ($b_t$) and lowest ask price ($a_t$). The bid-ask spread is defined as $s_t = a_t - b_t$. A sudden widening of $s_t$ indicates reduced liquidity or market stress.
    \item \textit{Order Book Volume ($V_{b}(t), V_{a}(t)$):} The total volume on the bid and ask sides of the order book. The market imbalance ratio $\rho_t$ is calculated as: $\rho_t=\frac{V_{b}(t) - V_{a}(t)}{V_{b}(t) + V_{a}(t)}$. High values of $\rho_t$ indicate directional pressure.
    \item \textit{Trade Volume ($v_t$):} The total quantity traded over a given period. Anomalous spikes are detected by examining deviations from the expected volume $\mu(v)$.
    \item \textit{Spread Width ($s_t$):} A key indicator of liquidity. A sharp increase in $s_t$ suggests a potential anomaly.
    \item \textit{Ladder Bid-Ask Depth ($D_b, D_a$):} The cumulative volume of orders at various price levels ($l$) away from the current price, calculated as $D_b=\sum_{i=1}^{l} v(b_i)$ and $D_a=\sum_{i=1}^{l} v(a_i)$. Significant changes in depth can indicate market stress.
    \item \textit{Inter-arrival Time ($\tau_t$):} The time difference between consecutive trades, $\tau_t=t_{i+1}-t_i$. Abnormal clustering may indicate high-frequency trading anomalies.
    \item \textit{Immediate Volatility ($\sigma_{\Delta p_{t}}$):} The standard deviation of price changes over a short window $\Delta t$. Abnormally high immediate volatility signals market stress \cite{MLPletteriStockTrading}\cite{femibLetteri24}.
    \item \textit{Realised Volatility ($\sigma_{R}$):} Measured over discrete intervals $T$ as $\sigma_{R}=\sqrt{\sum_{i=1}^{T}r_{i}^2}$, where $r_i=\log(p_i/p_{i-1})$ are the log-returns.
    \item \textit{Liquidity Index ($LQ_t$):} The Amihud Illiquidity Ratio \cite{COEN201923} is computed as $LQ_t=\frac{|r_t|}{v_t}$. A high $LQ_t$ signals poor liquidity.
\end{itemize}

\subsection{Outlier Detection Models}
\label{sec:models_list}
We evaluate a comprehensive suite of thirteen unsupervised models, grouped into statistical and machine learning categories.

\subsubsection{Statistical Models}
The initial approach for outlier detection relies on a suite of statistical models.

\textbf{Parametric Models:} Let \( X \in \mathbb{R}^{n \times p} \) be the dataset with \( n \) observations and \( p \) features. We assume the data follow a multivariate Gaussian distribution or are based on a robust estimate of the covariance.
    \begin{itemize}
        \item \textit{Elliptic Envelope (EE):} This method assumes the inlier data follows an elliptical distribution. It estimates a robust location $\boldsymbol{\mu}$ and covariance matrix $\boldsymbol{\Sigma}$. For each observation \( \mathbf{x}_i \in \mathbb{R}^p \), the Mahalanobis distance is calculated as:
        \begin{equation}
        d_M(\mathbf{x}_i) = \sqrt{(\mathbf{x}_i - \boldsymbol{\mu})^\top \boldsymbol{\Sigma}^{-1} (\mathbf{x}_i - \boldsymbol{\mu})}.
        \label{eq:mahal}
        \end{equation}
        An observation is classified as an outlier if its squared Mahalanobis distance exceeds a threshold derived from the chi-squared distribution:
   \begin{equation}\label{eq:bounds}
       \text{label}_i = \begin{cases}
       1 & \text{if } d_M(\mathbf{x}_i)^2 > \chi^2_{p, 1-\alpha} \quad (\text{outlier}), \\
       0 & \text{otherwise} \quad (\text{inlier}).
       \end{cases}
   \end{equation}
        \item \textit{Minimum Covariance Determinant (MCD):} The MCD method seeks the subset of \( h \) observations (where \( h = \lfloor n \cdot (1-\alpha) \rfloor \) and $\alpha$ is the expected outlier fraction) whose sample covariance matrix has the minimum determinant \cite{hubert_minimum_2010}.
$$(\boldsymbol{\mu}_{\text{MCD}}, \boldsymbol{\Sigma}_{\text{MCD}}) = \arg\min_{\substack{S \subseteq X: |S| = h}} \det(\boldsymbol{\Sigma}_S).$$
The Mahalanobis distance is then computed using these robust estimates, and outliers are identified using the same thresholding logic as in Equation \ref{eq:bounds}.
        \item \textit{Empirical Covariance (EC):} This method uses the classical sample mean $\bar{\mathbf{x}}$ and covariance matrix $\boldsymbol{\Sigma}$ to compute the Mahalanobis distance (Equation \ref{eq:mahal}). Outliers are identified by comparing the squared distance to a percentile of the resulting distance distribution (e.g., the 97.5th percentile).
    \end{itemize}

\textbf{Non-Parametric Models:} These methods do not require assumptions about the underlying data distribution.
    \begin{itemize}
        \item \textit{Histogram-Based Outlier Score (HBOS):} For each feature dimension, a histogram is constructed to estimate the density. The HBOS score for a data point \( \mathbf{x} \) is the sum of the inverse logarithms of the estimated densities for each of its feature values, assuming feature independence:
        $$\text{HBOS}(\mathbf{x}) = \sum_{i=1}^p \log\left(\frac{1}{\text{density}_i(x_i)}\right).$$
        Data points with scores exceeding a predefined threshold \( \tau \) are classified as outliers.
    \end{itemize}

\subsubsection{Unsupervised Machine Learning Models}
A range of unsupervised machine learning models is employed to detect more complex, non-linear patterns.

\begin{itemize}
    \item \textit{One-Class Support Vector Machine (OC-SVM):} OC-SVM learns a decision boundary that encompasses the majority of the data \cite{Nguyen2019MinhNghia}. The objective is to find a hyperplane that separates the data from the origin with maximum margin. The optimisation problem is:
    $$
    \min_{\mathbf{w}, \rho, \boldsymbol{\xi}} \frac{1}{2} \|\mathbf{w}\|^2 + \frac{1}{\nu n} \sum_{i=1}^n \xi_i - \rho
    $$
    subject to $ \mathbf{w} \cdot \phi(\mathbf{x}_i) \geq \rho - \xi_i, \quad \xi_i \geq 0 $, where $\nu \in (0, 1]$ is a hyperparameter that controls the fraction of outliers. We employ the Radial Basis Function (RBF) kernel.

    \item \textit{Density-Based Spatial Clustering of Applications with Noise (DBSCAN):} DBSCAN groups together points that are closely packed, marking as outliers those points that lie alone in low-density regions. It is governed by two parameters: $\epsilon$ (neighbourhood radius) and $\text{minPts}$ (minimum number of points to form a dense region).

    \item \textit{Isolation Forest (IsoF):} This ensemble method isolates observations by randomly selecting a feature and then randomly selecting a split value between the maximum and minimum values of that feature. The number of splits required to isolate a sample is averaged over a forest of trees to produce an anomaly score.

     \item \textit{Local Outlier Factor (LOF):} LOF measures the local density deviation of a data point with respect to its neighbours \cite{breunig_lof_2000}. The LOF score is computed as the ratio of the average local reachability density of its neighbours to its own local reachability density. A score significantly greater than 1 indicates an outlier.

     \item \textit{Clustering-Based Local Outlier Factor (CBLOF):} This method assigns an anomaly score to each data point based on its relationship with the cluster it belongs to. The score considers both the size of the cluster and the distance of the point to its cluster's centroid.

    \item \textit{K-Means:} This algorithm partitions data into \( k \) clusters. Points that are distant from any cluster centroid are considered potential outliers. The objective is to minimise the within-cluster sum of squares:
    \[
    J = \sum_{j=1}^k \sum_{\mathbf{x}_i \in C_j} \|\mathbf{x}_i - \boldsymbol{\mu}_j\|^2,
    \]
    where $\boldsymbol{\mu}_j$ is the centroid of cluster $C_j$.

    \item \textit{Ordering Points To Identify the Clustering Structure (OPTICS):} An extension of DBSCAN, OPTICS does not require specifying a global density parameter. It produces a reachability plot, from which clusters of varying densities can be extracted and outliers identified as points with high reachability distances.

    \item \textit{Subspace Outlier Detection (SOD):} SOD identifies outliers by finding them in various subspaces (subsets of features) of the data, which is particularly effective in high-dimensional settings.

    \item \textit{K-Nearest Neighbours (KNN):} In this context, the distance of a point to its \( k \)-th nearest neighbour is used as its outlier score. Large distances are indicative of anomalies.
\end{itemize}

For each model, hyperparameters were optimised via grid search and cross-validation, guided by the LOB application criteria outlined in Table \ref{tab:ml_modelsApplication}. For instance, DBSCAN's $\epsilon$ and $\text{minPts}$ were tuned to detect liquidity shocks, while OC-SVM's $\nu$ was optimised to minimise false positives in manipulation detection.

\begin{table*}[!ht]
\centering
\caption{Summary of Outlier Detection Models and Their Application in the LOB Context.}
\label{tab:ml_modelsApplication}
\begin{tabular}{llll}
\hline
\textbf{Category}    & \textbf{Models}                                                  & \textbf{Strengths}                                                              & \textbf{LOB Application}                                                   \\ \hline
Parametric           & EC, MCD, EE                                                      & \begin{tabular}[c]{@{}l@{}}Interpretability, \\ computational efficiency\end{tabular}   & \begin{tabular}[c]{@{}l@{}}Volume/spread \\ outliers\end{tabular}          \\ \hline
Non-Parametric       & HBOS                                                             & \begin{tabular}[c]{@{}l@{}}No distributional \\ assumptions\end{tabular}        & \begin{tabular}[c]{@{}l@{}}Temporal/liquidity \\ anomalies\end{tabular}    \\ \hline
Density-Based        & DBSCAN, OPTICS                                                   & \begin{tabular}[c]{@{}l@{}}Adapts to shifting \\ market conditions\end{tabular} & \begin{tabular}[c]{@{}l@{}}Liquidity shocks, \\ spoofing\end{tabular}      \\ \hline
Isolation/Proximity & \begin{tabular}[c]{@{}l@{}}Isolation Forest, \\ LOF, KNN\end{tabular} & \begin{tabular}[c]{@{}l@{}}Handles high-\\ dimensional data\end{tabular}        & \begin{tabular}[c]{@{}l@{}}Wash trading, \\ volatility spikes\end{tabular} \\ \hline
Clustering           & K-Means, CBLOF                                                   & \begin{tabular}[c]{@{}l@{}}Captures structural \\ market shifts\end{tabular}    & \begin{tabular}[c]{@{}l@{}}Layering, \\ artificial depth\end{tabular}      \\ \hline
\end{tabular}
\end{table*}

\subsection{From Outlier Score to Trading Signal: The AITA-OBS Pipeline}
The raw output of the unsupervised models is a numerical anomaly score. To translate this score into an actionable trading signal, we designed the AITA-OBS decision pipeline, a standardised process applied to each model to ensure a fair comparison.

\subsubsection{Anomaly Score Generation and Normalisation}
Let $M$ be one of the models under consideration. For each data point $X_t$, the model computes a raw anomaly score, $s_M^{\text{raw}}(X_t)$. As these scores are on different scales, we normalise them to a uniform range of $[0, 1]$ using Min-Max scaling over the entire dataset:
\begin{equation}
s_M(X_t) = \frac{s_M^{\text{raw}}(X_t) - \min(S_M^{\text{raw}})}{\max(S_M^{\text{raw}}) - \min(S_M^{\text{raw}})}
\label{eq:normalisation}
\end{equation}
where $S_M^{\text{raw}}$ is the set of all raw anomaly scores from model $M$.

\subsubsection{Binary Signal Generation via Dynamic Thresholding}
A normalised score is converted into a binary outlier signal, $\mathcal{O}_t \in \{0, 1\}$. We employ a dynamic threshold based on the historical distribution of scores. An observation $X_t$ is flagged as an outlier ($\mathcal{O}_t = 1$) if its score exceeds the 95th percentile of the score distribution for that model:
\begin{equation}
\mathcal{O}_t = 
\begin{cases} 
1 & \text{if } s_M(X_t) > \tau_M \\ 
0 & \text{otherwise} 
\end{cases}
\quad \text{where} \quad
\tau_M = P_{95}(\{s_M(X_t) \mid \forall t\})
\label{eq:binary_signal}
\end{equation}
The 95th percentile was chosen as a trade-off to filter market noise while capturing a sufficient number of significant events for trading.

\subsubsection{Trade Execution Logic and Backtesting Protocol}
Once a binary outlier signal is generated ($\mathcal{O}_t = 1$), a trade is executed.
\begin{itemize}
    \item \textbf{Trade Direction:} The strategy operates on a \textit{mean-reversion} assumption. The direction of the trade opposes the recent price momentum. If an outlier coincides with positive momentum, a \textbf{short} position is initiated; if it coincides with negative momentum, a \textbf{long} position is taken.
    \item \textbf{Position Sizing:} To ensure a fair comparison, a fixed fractional position sizing rule was applied \cite{vince_mathematics_1992}. For every trade, the amount invested was 33.33\% of the available capital at that time. This isolates the performance of the detection algorithm.
\end{itemize}
The backtesting simulation starts with an initial budget \( B_0 = \$1500 \). The profit for each trade is calculated based on the price change until the next signal or bar close. The cumulative profit ($P_T$) and percentage gain ($\%G$) are the final performance metrics. As a benchmark, we use a standard Buy-and-Hold (B\&H) strategy over the same period.

\section{Backtesting Results and Discussion}
\label{sec:results}
This section evaluates the performance of the statistical and machine learning models applied to the Bitcoin LOB data for outlier detection. The analysis leverages a dataset spanning 18 days (from 26 December 2024 to 13 January 2025), comprising 26,204 records at a 1-minute timeframe. Data preprocessing included normalisation, scaling, and feature engineering as described in Section \ref{sect:dataset}.

\begin{figure*}[ht]
    \centering
    \includegraphics[width=1.05\linewidth]{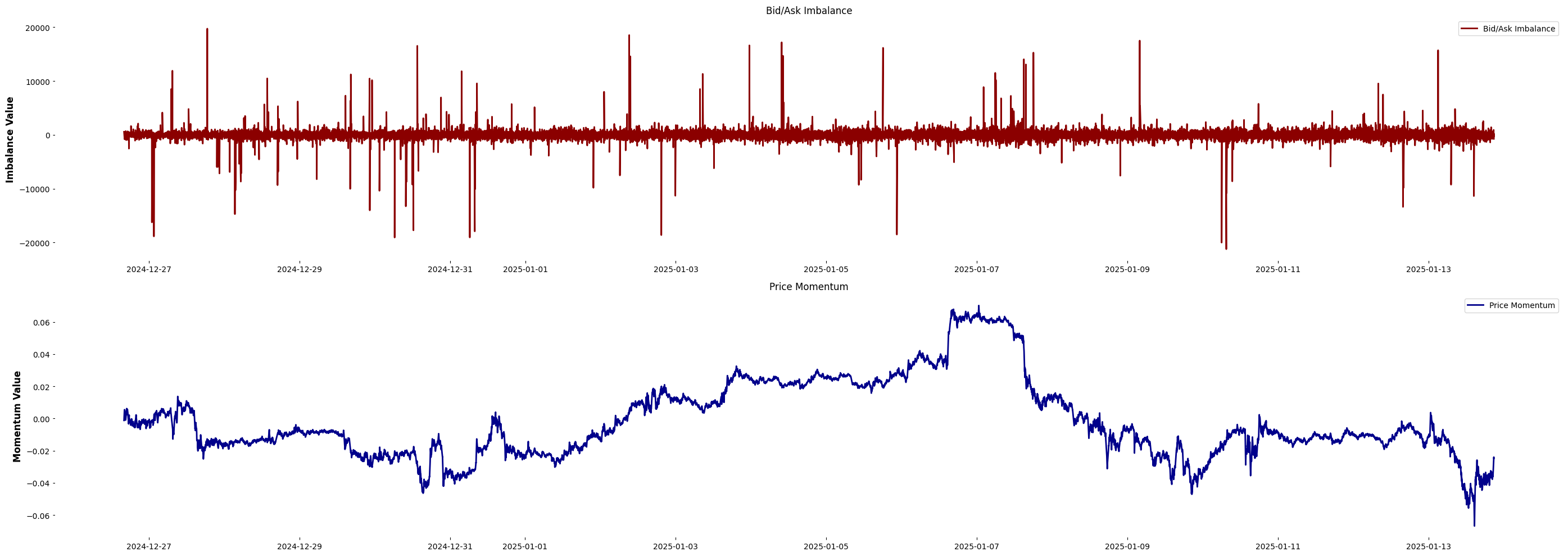}
    \caption{The bid/ask imbalance variability and the price momentum over the evaluation period.}
    \label{fig:outputImbMom}
\end{figure*}

The \textit{Bid/Ask Imbalance} and \textit{Price Momentum} plots in Figure \ref{fig:outputImbMom} confirm the expected dynamics of order flow and price changes over the trading period. The bid/ask imbalance exhibits high variability, particularly during periods of heightened volatility, while the price momentum graph highlights the strategy's ability to capitalise on sustained price trends.

The backtesting strategy was implemented as follows: Let \( B_0 = \$1500 \) be the initial budget. Let \( B_t \) be the budget, \( P_t \) the cumulative profit, and \( R_t \) the return at time \( t \). Initially, \( B_t = B_0 \), \( P_t = 0 \), and \( R_t = 0 \). For each time step \( t \), if an outlier is detected, a trade (long or short) is executed: (i) Trade Amount = $0.3333 \times B_{t-1}$; (ii) Price Change $\Delta\% = (\text{Close}(t-1) - \text{Close}(t)) / \text{Close}(t-1)$; (iii) Profit = Amount $\times \Delta\%$; (iv) $B_t = B_{t-1} + \text{Profit}$; (v) $P_t = B_t - B_0$; (vi) $R_t = \text{Profit} / B_0$. The win rate (\( \text{WR} \)) is calculated as \( \text{WR} = (\text{Winning Trades} / \text{Total Trades}) \times 100 \). As a benchmark, the B\&H profit is calculated as $\text{B\&H Profit} = B_0 \times (1 + \sum_{t=1}^{T} \text{DR}(t)) - B_0$, where \( \text{DR}(t) \) is the daily return. The final results include the cumulative profit (\( P_T \)) and the percentage gain (\( \%G \)): $P_T = B_T - B_0$ and $\%G = (P_T / B_0) \times 100$.

\subsection{Statistical Model Performance}
\begin{table*}[!ht]
\centering
\caption{Performance of Statistical Models.}
\label{tab:stat_models}
\sisetup{table-format=-1.2, table-number-alignment=center}
\begin{tabular}{l S[table-format=3.0] S[table-format=3.0] S[table-format=3.2] S[table-format=2.2]}
\toprule
\textbf{Model} & {\textbf{\# Long Trades}} & {\textbf{\# Short Trades}} & {\textbf{Cum. Profit (\$)}} & {\textbf{Gain (\%)}} \\ 
\midrule
EE             & 63                      & 69                       & -8.23                           & -0.55                      \\
HBOS           & 562                     & 723                      & 70.27                           & 4.68                       \\
MCD            & 347                     & 309                      & -8.89                           & -0.59                      \\
\textbf{EC}    & \textbf{341}            & \textbf{315}             & \textbf{100.47}                 & \textbf{6.70}              \\ 
\bottomrule
\end{tabular}
\end{table*}

\begin{figure*}[!ht]
    \centering
    \includegraphics[width=\textwidth]{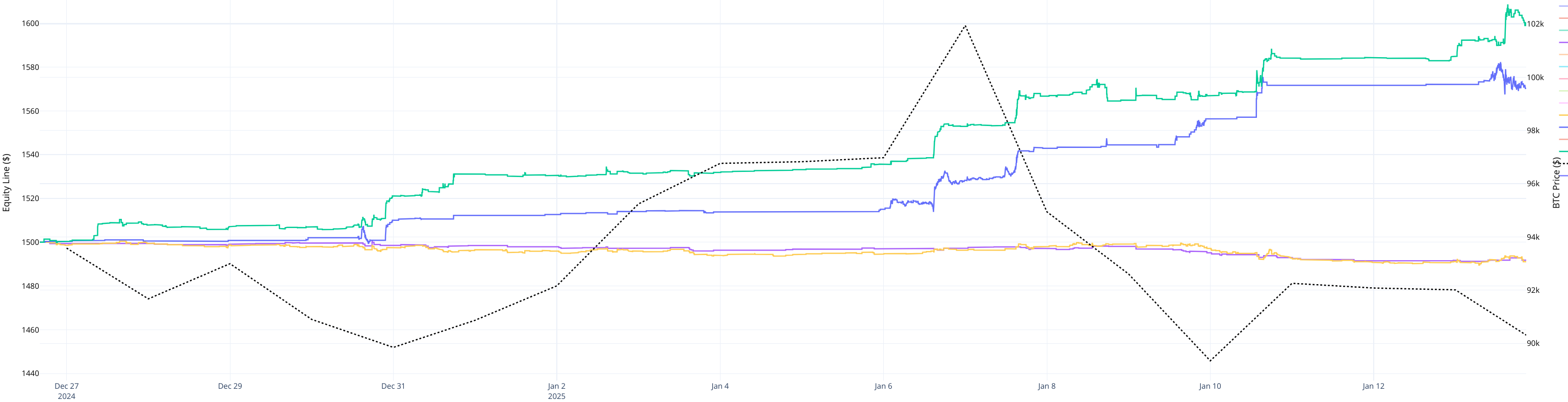}
    \caption{Equity curves of statistical models. The dotted black line represents the daily B\&H benchmark.}
    \label{fig:equity_stats}
\end{figure*}

Table \ref{tab:stat_models} shows that Empirical Covariance (EC) was the most effective statistical model, achieving a cumulative profit of \$100.47 (a 6.70\% gain). Its equity curve (Figure \ref{fig:equity_stats}) showed consistent growth. HBOS also performed well, with a \$70.27 profit, but its high trade frequency suggests higher potential transaction costs. In contrast, MCD and EE underperformed, incurring small losses, indicating their limitations in this context.

\subsection{Machine Learning Model Performance}
\begin{table*}[!ht]
\centering
\caption{Performance of Machine Learning Models.}
\label{tab:ml_model}
\sisetup{table-format=-1.2, table-number-alignment=center}
\begin{tabular}{l S[table-format=4.0] S[table-format=4.0] S[table-format=2.2] S[table-format=1.2]}
\toprule
\textbf{Model} & {\textbf{\# Long Trades}} & {\textbf{\# Short Trades}} & {\textbf{Cum. Profit (\$)}} & {\textbf{Gain (\%)}} \\ 
\midrule
\textbf{CBLOF}           & \textbf{649}              & \textbf{662}              & \textbf{75.48}                  & \textbf{5.03}               \\
OPTICS                   & 19                        & 20                        & 12.86                           & 0.86               \\
K-Means                  & 4354                      & 4498                      & 1.10                            & 0.07               \\
KNN                      & 674                       & 637                       & 71.79                           & 4.79               \\
\textbf{OC-SVM}          & \textbf{71}               & \textbf{69}               & \textbf{43.59}                  & \textbf{2.91}               \\
SOD                      & 1283                      & 1338                      & 42.81                           & 2.85               \\
LOF                      & 127                       & 136                       & 60.34                           & 4.02               \\
IsoF                     & 135                       & 128                       & 12.84                           & 0.86               \\
DBSCAN                   & 87                        & 89                        & 32.42                           & 2.16               \\ 
\bottomrule
\end{tabular}
\end{table*}

\begin{figure*}[!ht]
    \centering
    \includegraphics[width=\textwidth]{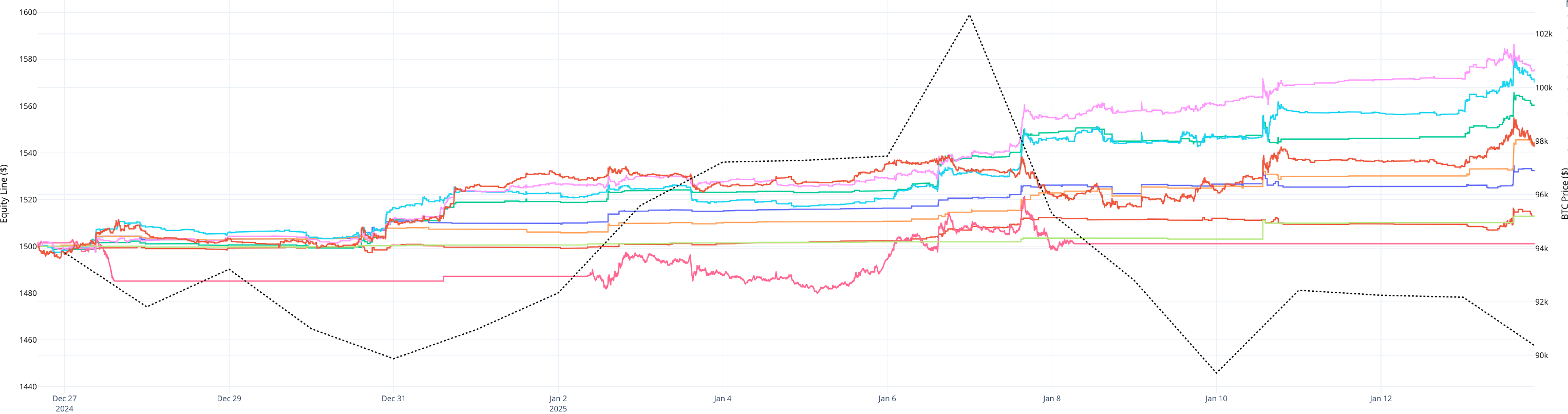}
    \caption{Equity curves of machine learning models. The dotted black line represents the daily B\&H benchmark.}
    \label{fig:equity_ml}
\end{figure*}

Among the machine learning models (Table \ref{tab:ml_model}), CBLOF yielded the highest cumulative profit at \$75.48 (a 5.03\% gain), though it executed a high number of trades. KNN performed similarly. Notably, OC-SVM provided a respectable gain of \$43.59 (2.91\%) with only 140 trades, suggesting a more efficient strategy when considering transaction costs. The equity curves in Figure \ref{fig:equity_ml} depict these varied performances.

\subsection{Benchmark Comparison and Cost Analysis}
\begin{figure}[!ht]
    \centering
    \includegraphics[width=0.7\linewidth]{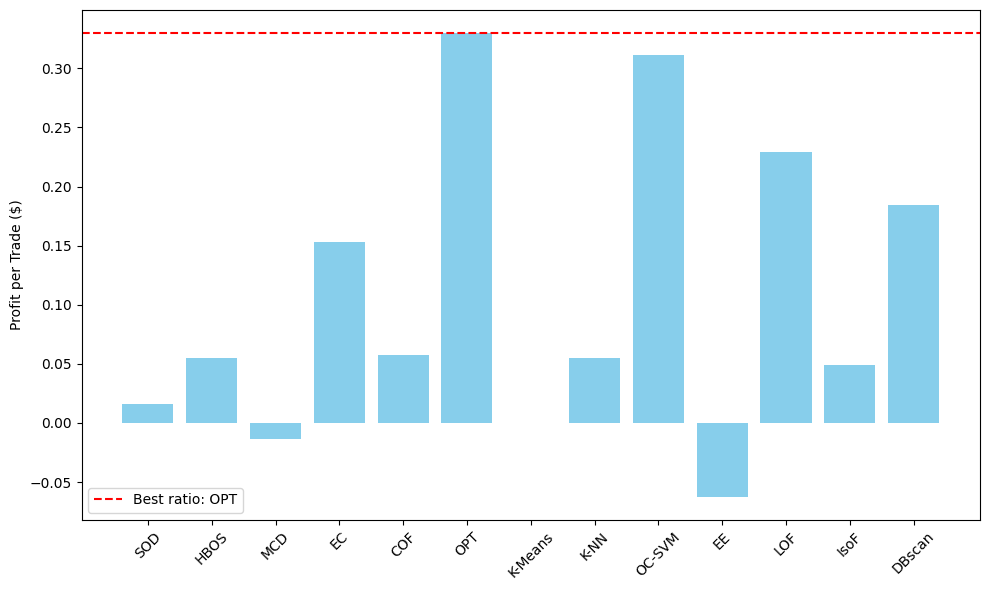}
    \caption{Ratio of cumulative profit per trade for each model.}
    \label{fig:bestRatio}
\end{figure}

The statistical EC model was the top performer overall. The best machine learning model, CBLOF, achieved a strong gain but with nearly double EC's number of trades, highlighting the critical importance of transaction costs. To assess trade efficiency, we calculated the cumulative profit per trade (Figure \ref{fig:bestRatio}). Models like OC-SVM and OPTICS demonstrate a superior profit-to-trade ratio, making them potentially more viable in real-world scenarios where fees are a significant factor.

Crucially, with the exception of the EE and MCD models, all other strategies generated positive returns. The top-performing models, in particular, significantly outperformed the B\&H benchmark, which incurred a loss of -\$37.06 over the same period. This demonstrates the general effectiveness of the proposed outlier-driven, mean-reversion trading strategy.

\begin{figure}[!ht]
    \centering
    \includegraphics[width=0.7\linewidth]{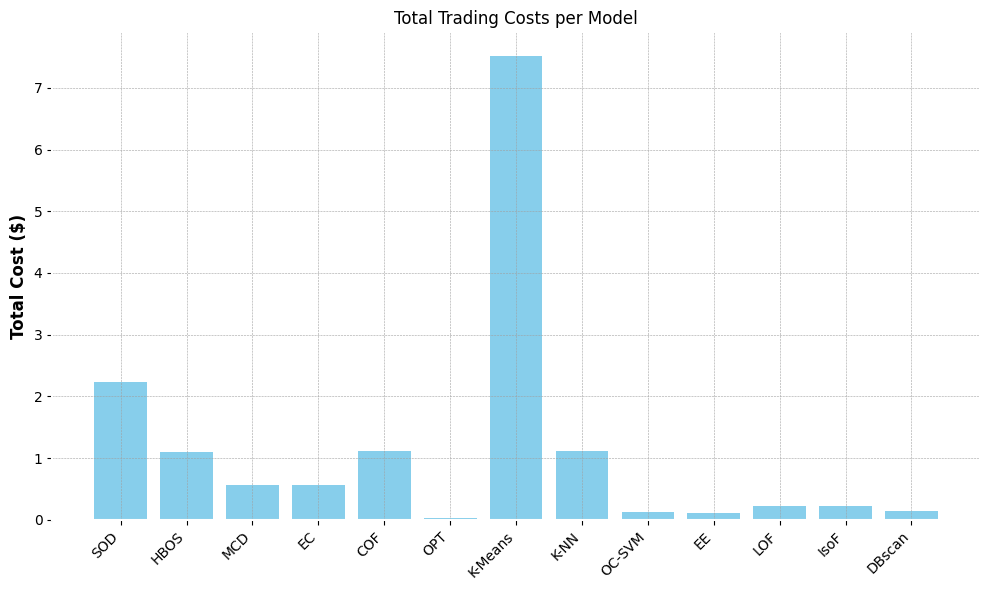}
    \caption{Estimated total cost of fees per trade for each model.}
    \label{fig:costs}
\end{figure}

Figure \ref{fig:costs} illustrates the estimated total trading fees for each model, assuming a market maker fee of 0.08\% per trade. The high trade frequency of models like K-Means, CBLOF, and KNN results in substantial costs that would erode profits. In contrast, the EC and OC-SVM models incur significantly lower costs, reinforcing their practical viability.

\section{Conclusion}
\label{sec:conclusion}

This study presented a rigorous comparative analysis of thirteen statistical and machine learning models for anomaly detection in Bitcoin LOBs. By employing a unified testing environment, AITA-OBS, we provided a transparent and reproducible evaluation of each model's ability to generate profitable trading signals based on a mean-reversion strategy.

Our empirical evaluation demonstrates that simple, robust statistical methods can be highly effective. The Empirical Covariance (EC) model emerged as the top-performing strategy, achieving a 6.70\% gain with a moderate number of trades. Among the machine learning techniques, while models like CBLOF showed high profitability, their high trade frequency raises concerns about transaction costs in real-world applications. In contrast, the OC-SVM model offered a compelling balance between profitability and trade efficiency, making it a strong candidate for cost-sensitive strategies. The analysis confirms that most outlier detection models, when coupled with a simple contrarian logic, can significantly outperform a standard Buy-and-Hold benchmark in a volatile market.

This work provides a valuable benchmark for researchers and practitioners. Future research should focus on dynamic thresholding mechanisms and the integration of the best-performing models into adaptive ensemble systems. Furthermore, the framework's adaptability could be tested on a wider range of digital assets and traditional markets to evaluate its generalisability. In summary, our findings highlight the significant potential of outlier-driven strategies for risk management and alpha generation in complex financial markets.

In near future work, we aim to enhance optimisation strategies \cite{letteri2020DOS},\cite{LetteriCP2021intellisys}, and we will expose the AITA framework's API as a secure service to thwart botnet attacks using Deep Learning models \cite{LetteriPG19SecIOT}\cite{LetteriPC19HTTP}. We plan to create a Multi-agent System which features transparent Ethical Agents for customer service \cite{DyoubCLL20} or combines logic constraint and DRL \cite{GasperisCRMLD23}. We will evaluate dialogues with the AITA agents \cite{Letteri2109ethMon} with guidance from an ethical teacher \cite{DyoubCL22ethTeach} due to the critical field combining all results with related contexts like technology-enhanced learning \cite{mis4tel2023}\cite{mis4telLetteriV24}. 

\bibliographystyle{apalike}  
\bibliography{bibliography}

\end{document}